\documentclass[prd,showpacs,preprintnumbers, aps]{revtex4-1}
\usepackage[applemac]{inputenc}
\usepackage[T1]{fontenc}
\def\beq{\begin{equation}}
\def\eeq{\end{equation}}
\def\bea{\begin{eqnarray}}
\def\eea{\end{eqnarray}}
\def\ba{\begin{array}}
\def\ea{\end{array}}

\def\part{\partial}

\begin{document}
\preprint{UdeM-GPP-TH-08-173}
\preprint{arXiv:0812.2491[gr-qc]}
\title{Monochromatic plane-fronted waves in conformal gravity are pure gauge}
\author{Luca Fabbri$^{1,2}$}
\email{luca.fabbri@bo.infn.it}
\author{M.~B.~Paranjape$^{1}$}
\email{paranj@lps.umontreal.ca}
\affiliation{$^1$Groupe de physique des particules, D\'epartement de physique,\\
Universit\'e de Montr\'eal, C.P. 6128, succ. Centre-ville, Montr\'eal, Qu\'ebec, Canada H3C 3J7}
\affiliation{$^2$Theory Group, I.N.F.N. \& Department of Physics,\\
University of Bologna, Via Irnerio 46, C.A.P. 40126, Bologna, Italy}
\begin{abstract}
We consider  plane fronted, monochromatic gravitational waves on a Minkowski background, in a conformally invariant theory of general relativity.  By this we mean waves of the form:
$g^{\mu\nu}=\eta^{\mu\nu} +\epsilon^{\mu\nu} F(k\cdot x)$, where $\epsilon^{\mu\nu}$ is a constant polarization tensor, and $k_\mu$ is a light like vector.  We also assume the coordinate gauge condition $|g|^{-1/4}\partial_\tau(|g|^{1/4}g^{\sigma\tau})=0$ which is the conformal analog of the harmonic gauge condition $g^{\mu\nu}\Gamma^\sigma_{\mu\nu}=-|g|^{-1/2}\partial_\tau(|g|^{1/2}g^{\sigma\tau})=0$, where $\det\left[g_{\mu\nu}\right]\equiv g$.  Requiring additionally the conformal gauge condition $ g=-1$  surprisingly implies that the waves are both transverse and traceless.  Although the ansatz for the metric is eminently reasonable when considering perturbative gravitational waves, we show that the metric  is reducible to the metric of Minkowski space-time via a sequence of  coordinate transformations which respect the gauge conditions, without any perturbative approximation that $\epsilon^{\mu\nu}$ be small.   This implies that we have in fact, exact plane wave solutions, however they are simply coordinate/conformal artifacts. As a consequence, they carry no energy.  Our result does not imply that conformal gravity does not have gravitational wave phenomena.  A different, more generalized ansatz for the deviation, taking into account the fourth order nature of the field equation, which has the form  $g^{\mu\nu}=\eta^{\mu\nu} +B^{\mu\nu}(n\cdot x) G(k\cdot x)$ indeed yields waves which carry energy and momentum, \cite{m}.  It is just surprising that transverse, traceless, plane fronted gravitational waves, those that would be used in any standard, perturbative, quantum analysis of the theory, simply do not exist.   
\end{abstract}
\pacs{04.20.-q, 04.20.Cv, 04.20.Jb, 04.30.Nk, 04.50.-h}
\pagenumbering{arabic}
\maketitle
\section{Introduction}
Einstein gravity explains most of the observed gravitational phenomena in the universe, however, a few observations appear to lie outside of the reach of its explanatory power without a drastic modification of the matter content. Since such a modification is not reasonable for luminous matter, as observational data greatly constrain the possibility to tinker with the distribution and quantity of luminous matter, Einsteinian relativity is salvaged via the addition of copious quantities of dark matter and energy. This situation is epistemologically unsatisfactory and some effort has been made to investigate alternative theories of relativity. The geometrical structure of Einsteinian relativity can be enriched by imposing the principle of scale invariance, and this in fact implies a unique gravitational action in four dimensions: the corresponding theory is called Weyl gravity. Weyl gravity can be used to explain the remaining observed phenomena - those not well described by Einstein gravity and the existing spectrum of luminous matter (see for example \cite{m1-k}, \cite{m2}, \cite{m3}).

Gravitational waves are inescapable predictions of Einsteinian relativity, and their indirect observation has already been rewarded with a Nobel prize \cite{h-t}, although their direct detection is still a decade away in principle. Of course, when entertaining alternative theories of gravity, one would then be very interested in elaborating the existence of such waves. Here we begin such an investigation for Weyl gravity.
\section{Weyl gravitational plane waves}
\subsection{Weyl gravity}
The general structure of Weyl gravity is given by a metric $g_{\mu\nu}(x)$ which is a coordinate tensor and which transforms as $g^{c}_{\mu\nu}(x)=\Omega^{2}(x)g_{\mu\nu}(x)$ for a conformal transformation specified by $\Omega(x)$. The metric gives rise to the Levi-Civita connection $\Gamma^{\rho}_{\mu\nu}(x)$ the Riemann tensor $R_{\alpha\mu\sigma\rho}(x)$ and the Weyl conformal tensor $C_{\alpha\mu\sigma\rho}(x)$ (for the notation throughout this paper we will follow \cite{e-f1-p}, see also \cite{f2}). The importance of Weyl conformal tensor lies in the fact that it is covariant under both coordinate and scale transformations; moreover, in four and higher dimensional space-times, it vanishes if and only if there exists a combination of these two transformations for which the metric $g_{\mu\nu}$ can be brought to $\eta_{\mu\nu}$ the metric of flat Minkowski space-time.

The Riemann tensor is not conformally covariant, hence it contains gauge artifacts that are removable by conformal transformations. The Weyl conformal tensor on the other hand is conformally covariant, $C_{\alpha\mu\sigma\rho}(x)\to \Omega(x)^2C_{\alpha\mu\sigma\rho}(x)$. Hence it is the fundamental object that contains all information about the geometrical background. In four dimensional geometries the dynamics is uniquely determined by
\beq
{\cal L}=C_{\alpha\mu\sigma\rho}C^{\alpha\mu\sigma\rho}+k{\cal L}^{\mathrm{matter}}
\eeq
which gives rise to field equations of the form
\begin{equation}
W_{\mu\nu}=-\frac{k}{4}T_{\mu\nu}^{\mathrm{matter}}
\end{equation}
where $W_{\mu\nu}$ is the Bach conformal tensor
\beq
W_{\mu\nu}=
\nabla^{2}R_{\mu\nu}-\frac{1}{3}\nabla_{\mu}\nabla_{\nu}R-\frac{1}{6}g_{\mu\nu}\nabla^{2}R+
2R^{\alpha\beta}(R_{\beta\nu\alpha\mu}-\frac{1}{4}g_{\mu\nu}R_{\alpha\beta})
-\frac{2}{3}R(R_{\mu\nu}-\frac{1}{4}g_{\mu\nu}R)
\eeq
and $k$ is a constant related to the gravitational constant and $T_{\mu\nu}^{\mathrm{matter}}$ is the energy-momentum tensor of the matter fields.
\subsection{Transverse, traceless, plane fronted, monochromatic plane waves}
To investigate wave solutions of a set of field equations, the usual approach followed in Einstein gravity consists of taking a linear perturbation of the form
\begin{equation}
g^{\mu\nu}=\eta^{\mu\nu}+\epsilon^{\mu\nu} F(k \cdot x)\label{1}
\end{equation}
where $k \cdot x \equiv k_{\mu}x^{\mu}=\omega (t-x)$, $\omega$ a constant frequency, and where the propagation vector is to be light-like (see for example \cite{w}). The perturbing term is assumed to be small when considering the linearized theory, usually higher order corrections are not examined. Clearly, it is of interest to consider such perturbations in Weyl gravity. Infact, we will find that Eq. (\ref{1}), taking into account the restrictions implied by assuming appropriate gauge fixing, is actually an exact, albeit trivial, solution to the full non-linear field equations with no assumption of perturbation.

We must impose two gauge conditions, one fixing coordinate transformations while the other fixing conformal rescalings.  One possibiity is to  impose, where as usual $g\equiv \det[g_{\mu\nu}]$, 
\begin{equation}
\Gamma^{\rho} \equiv \Gamma^{\rho}_{\mu\nu}g^{\mu\nu}=
-{{\vert g \vert}^{-1/2}}\partial_{\mu}({\vert g \vert}^{1/2}g^{\mu\rho})=0\label{w}
\end{equation}
which is called the harmonic gauge \cite{w}, essentially fixing coordinate transformations (up to harmonic coordinate transformations).  Secondly we would impose the conformal gauge condition
\beq
g=-1
\eeq
which fixes the possibility to perform any conformal transformations.  However, the two conditions are not necessarily compatible.  Performing a conformal transformation to impose the conformal gauge  in general destroys the harmonic gauge condition.  We could simply, tacitly assume that the metric is such that both are satisfied.  However, instead, we impose\cite{th}

\beq
-\frac{1}{{\vert g \vert}^{1/4}}\partial_{\mu}(\vert g \vert^{1/4}g^{\mu\rho})=0.\label{gc}
\eeq
This condition was identified in Ref. \cite{mcgc} as an evidently conformally invariant alternative to the usual harmonic gauge condition.  It also allows the possibility of performing a restricted class of coordinate transformations, analogous to  harmonic coordinate transformations, which preserve it.  We will also impose the conformal gauge choice $g=-1$, which then, in conjunction with Eqn. (\ref{gc}) simply gives
\beq
\partial_\mu g^{\mu\rho}=0.\label{dz}
\eeq
Of course, the conformal gauge condition in conjunction with the harmonic gauge condition Eqn. (\ref{w}), evidently gives the same condition Eqn. (\ref{dz}), however, as stated above, it is not in general possible to perform the necessary conformal transformation without ruining the harmonic gauge, or vice versa.

We will impose the conformal gauge condition only after analyzing the coordinate gauge condition Eqn.(\ref{gc}), which  gives explicitly
\beq
\epsilon^{\rho\nu}k_{\nu}(F|g|^{1/4})'+k^{\rho}(|g|^{1/4})'=0\label{hg}
\eeq
where the $'$ means differentiation with respect to the argument, and due to the ansatz for the metric, its determinant $g=g(k\cdot x)$ is necessarily a function of $k\cdot x$ and we use the definition $k^\rho\equiv\eta^{\rho\nu}k_\nu$.   Writing out the equations that are implied and using $k_0=-k_1=\omega=k^0=k^1$, we get:

\bea
(\epsilon^{00}- \epsilon^{01})(F|g|^{1/4})'+(|g|^{1/4})'&=&0\label{h1}\\
(\epsilon^{10}- \epsilon^{11})(F|g|^{1/4})'+(|g|^{1/4})'&=&0\label{h2}\\
(\epsilon^{20}- \epsilon^{21})(F|g|^{1/4})'&=&0\\
(\epsilon^{30}- \epsilon^{31})(F|g|^{1/4})'&=&0
\eea
We can reject the solution to the latter two equations $(F|g|^{1/4})'=0$, since when inserted in the first two equations it implies   $(|g|^{1/4})'=0$.    Put back in  $(F|g|^{1/4})'=0$ then gives $F=const.$ which is not a wave.  Thus we require:
\bea
(\epsilon^{20}- \epsilon^{21})&=&0\\
(\epsilon^{30}- \epsilon^{31})&=&0
\eea
Now spelling out Eqns. (\ref{h1}) and (\ref{h2}) gives:
\bea
(\epsilon^{00}- \epsilon^{01})&=&-(|g|^{1/4})'/(F|g|^{1/4})'\\
(\epsilon^{10}- \epsilon^{11})&=& -(|g|^{1/4})'/(F|g|^{1/4})'
\eea
Since the right hand sides are in principle functions of $k\cdot x$ these equations make no sense except if 
\beq
-(|g|^{1/4})'/(F|g|^{1/4})'=D\label{constraint},
\eeq
with $D$ a constant.  This integrates simply as
\beq
|g|^{1/4}=\frac{C}{1+DF}\label{int}
\eeq
with $C$ an integration constant.  Actually we will choose $D=0, \,\, C=1$ by a conformal transformation rendering the determinant of the metric equal to -1,  but we will carry them along for the moment.  Thus we get the solution:
\bea
\epsilon^{00}&\equiv&\epsilon\\
\epsilon^{01}=\epsilon^{10}&=&\epsilon-D\\
\epsilon^{11}&=& \epsilon-2D\\
\epsilon^{20}=\epsilon^{02}= \epsilon^{21}=\epsilon^{12}&=&b\\
\epsilon^{30}=\epsilon^{03}=\epsilon^{31}=\epsilon^{13}&=&c
\eea
and the metric ansatz must have the form
\begin{equation}
\left[g^{\mu\nu}\right]=\left[\eta^{\mu\nu}\right]+\left[\epsilon^{\mu\nu}\right]F=
\left[\eta^{\mu\nu}\right]+\left(\begin{tabular}{cccc}
$\epsilon $ & $\epsilon -D $ & $ b $ & $ c $\\
$\epsilon-D $ & $\epsilon-2D $ & $ b $ & $ c $\\ 
$ b $ & $ b $ & $ -D+f $& $ j $\\
$ c $ & $ c $& $ j $ & $ -D+h $
\end{tabular}\right)F.
\end{equation}
The constants $b$ and $c$ are completely arbitrary, while the constants $\epsilon, f, h, j, D$ are also arbitrary up to the limits due to the constraint implied by Eqn. (\ref{constraint}).  The metric can be more obviously written as
\begin{equation}
\left[g^{\mu\nu}\right]=\left[\eta^{\mu\nu}\right]+\left[\epsilon^{\mu\nu}\right]F=
(1+DF)\left[\eta^{\mu\nu}\right]+\left(\begin{tabular}{cccc}
$\epsilon-D $ & $\epsilon -D $ & $ b $ & $ c $\\
$\epsilon-D $ & $\epsilon-D $ & $ b $ & $ c $\\ 
$ b $ & $ b $ & $ f $& $ j $\\
$ c $ & $ c $& $j $ & $ h $
\end{tabular}\right)F
\end{equation}
and it is easy to calculate its determinant exactly.  We get
\beq
1/g=(1+DF)^2(-(1+DF)^2+(1+DF)(f+h)F+(j^2-fh)F^2).\label{det}
\eeq
 and using this in Eqn. (\ref{int}), assuming $g<0$, gives
\beq
(-(1+DF)^2+(1+DF)(f+h)F+(j^2-fh)F^2)=-\frac{(1+DF)^2}{C^4}.
\eeq
We can write this equation in the form
\beq
\alpha (1+DF)^2 +\beta (1+DF)F +\gamma F^2=0\label{eqc}
\eeq
with $\alpha=-1+1/C^4$,  $\beta=f+h$ and $\gamma=j^2-fh$.  
If we expand out the polynomial in $F$ we find the constant term is simply $\alpha$,  hence we get $\alpha=0$.  This means
\beq
\alpha = -1+\frac{1}{C^4}=0
\eeq
{\it i.e.} $C=1$.  But then looking at the linear term that remains gives $\beta=0$, and  we get for the $F^2$ term $\beta D+\gamma=0$ which yields $\gamma=0$.  Thus $f=-h$ and then $j^2+f^2=0$, which requires that $j=0$ and $f=h=0$.  However $D$ is completely arbitrary.  Thus we find 
\begin{equation}
\left[g^{\mu\nu}\right]=\left[\eta^{\mu\nu}\right]+\left[\epsilon^{\mu\nu}\right]F=
(1+DF)\left[\eta^{\mu\nu}\right]+\left(\begin{tabular}{cccc}
$\epsilon-D $ & $\epsilon -D $ & $ b $ & $ c $\\
$\epsilon-D $ & $\epsilon-D $ & $ b $ & $ c $\\ 
$ b $ & $ b $ & $ 0 $& $ 0$\\
$ c $ & $ c $& $0 $ & $ 0$
\end{tabular}\right)F\label{metric1}
\end{equation}
The trace of the metric, $\eta_{\mu\nu}g^{\mu\nu}=4(1+DF)=4+\eta_{\mu\nu}\epsilon^{\mu\nu}F$.   This is as far as we can go without imposing a choice of the conformal gauge.  We choose the conformal gauge to be 
\beq
g=-1\label{2}.
\eeq
The determinant of the metric in Eqn. (\ref{metric1}) can be easily computed, we find  $1/g=-(1+DF)^4$.  Thus imposing the conformal gauge condition Eqn. (\ref{2}) fixes $D=0$ and which simultaneously makes the polarization tensor trace free, $\eta_{\mu\nu}\epsilon^{\mu\nu}=0$.  The coordinate gauge condition then also makes the polarization  tensor transverse, directly from Eqn. (\ref{hg}), we get, $\epsilon^{\rho\mu}k_\mu (F|g|^{1/4})'=0$ {\it ie.} $\epsilon^{\rho\mu}k_\mu =0$.  Thus we can choose the coordinate and conformal gauge conditions so that the metric deviation is transverse and traceless.

%
Thus we find the metric ansatz is reduced to 
\begin{equation}
\left[g^{\mu\nu}\right]=\left[\eta^{\mu\nu}\right]+\left[\epsilon^{\mu\nu}\right]F=
\left[\eta^{\mu\nu}\right]+\left(\begin{tabular}{cccc}
$\epsilon $ & $\epsilon $ & $ b $ & $ c $\\
$\epsilon$ & $\epsilon$ & $ b $ & $ c $\\ 
$ b $ & $ b $ & $ 0$& $ 0 $\\
$ c $ & $ c $& $0 $ & $ 0 $\label{fansatz}
\end{tabular}\right)F
\end{equation}
when the coordinate gauge  Eqn. (\ref{gc}) and the conformal gauge $g=-1$ are imposed.  This form is obviously transverse and traceless.

This result is valid exactly, no perturbative assumption was made.  To reiterate, the only assumptions made were the form for the metric ansatz Eqn. (\ref{1}) and that it satisfies the coordinate gauge condition Eqn. (\ref{gc}) and the particular choice of conformal gauge Eqn. (\ref{2}).   These then imposed the final form for the polarization tensor in Eqn. (\ref{fansatz}).    

If we examine the perturbative description a little, and write each component as an expansion in a small parameter, for example $f=f^{(1)}+f^{(2)}+\cdots$
where $f^{(i)}$ is of the $i$th order in the small parameter, we find Eqn. (\ref{eqc}) yields with $D=0$ 
\beq
f^{(1)}+h^{(1)}=0\\
\eeq
and 
\beq
f^{(2)}+h^{(2)}=0, \quad\quad  -f^{(1)}h^{(1)}+(j^{(1)})^2=0 \quad\quad\cdots
\eeq
Thus to second order we find
\beq
(f^{(1)})^2+(j^{(1)})^2=0, 
\eeq
implying $f^{(1)}=j^{(1)}=0$, which are the usual transverse, tracefree conditions obtained perturbatively. 

Taking the form of the metric given then in Eq. (\ref{fansatz}), with a little calculation we find, in fact for any $F=F(kx)$ a generic differentiable function,
\begin{equation}
g_{\mu\nu}=\eta_{\mu\nu}-\epsilon_{\mu\nu}F-(b^{2}+c^{2})k_{\mu}k_{\nu}F^{2}
\label{metric}
\end{equation}
where
\begin{equation}
\epsilon_{\mu\nu}=\eta_{\mu\sigma}\eta_{\nu\tau}\epsilon^{\sigma\tau}.
\end{equation}

As mentioned above, the coordinate gauge condition (\ref{gc}) does not completely fix the possible choice of coordinates; there exist coordinate transformations that preserve this gauge choice. Harmonic coordinate transformations of the form
\beq
x^{\prime\mu}=x^\mu +\epsilon^\mu\Phi(kx)
\eeq
where $\Phi(y)=\int^y dz F(z)$, the indefinite integral of $F$, we get
\begin{equation}
\frac{\partial x'^{\mu}}{\partial x^{\nu}}=\delta^{\mu}_{\nu}+\epsilon^{\mu}k_{\nu}F,
\end{equation}
which is easily seen to preserve the metric Eqn. (\ref{1}) ansatz and the harmonic gauge Eqn. (\ref{w}).  But the coordinate gauge choice Eqn. (\ref{gc}) nor the conformal gauge Eqn. (\ref{2}) are then, in general, not respected. However, with the choice of $\epsilon^\mu$ given by
\begin{equation}
\left[\epsilon^{\mu}\right]=\left(
\begin{tabular}{c}
A \\
A \\ 
B \\
C
\end{tabular}\right)
\end{equation}
it is easily verified that both the coordinate gauge condition Eqn. (\ref{gc}) and the conformal gauge condition Eqn. (\ref{2}) are respected. The crucial point of this paper is that the specific choice
\begin{equation}
\left[\epsilon^{\mu}\right]=-\frac{1}{2}\left(
\begin{tabular}{c}
$\epsilon$ \\
$\epsilon$ \\ 
$2b$ \\
$2c$
\end{tabular}\right)\label{5}
\end{equation}
brings the metric given in $(\ref{1})$ and $(\ref{metric})$ to $\eta_{\mu\nu}$, the flat Minkowskian form.

Thus, monochromatic plane-fronted wave fluctuations of the metric about flat Minkowski space-time are indeed solutions of the exact Weyl field equations. But they are trivial solutions: they are gauge (coordinate and conformal) equivalent to the vacuum itself, and nothing more. In particular, if the linearized approximation to Weyl conformal gravity makes sense, then this approximation simply does not admit gravitational wave phenomena of the type considered. Clearly a linearized approximation admits the ansatz (\ref{1}), the general linearized ansatz being a linear superposition of such waves. Again, linearized superpositions of harmonic coordinate transformations (\ref{5}), will bring the metric to the flat metric. Thus gravitational waves in Weyl gravity could only possibly exist in non-linear superpositions, where specifically the non-linear terms are not neglected. 

A simple calculation of the connection gives the result
\begin{equation}
\Gamma^{\mu}_{\alpha\beta}=
-\frac{1}{2}F'(k_{\alpha}\epsilon^{\mu}_{\beta}
+k_{\beta}\epsilon^{\mu}_{\alpha}
-k^{\mu}\epsilon_{\alpha\beta})
\end{equation}
which vanishes by the harmonic transformations above and its Riemann tensor is given by
\begin{equation}
R_{\mu\alpha\rho\beta}=
-\frac{1}{2}F''(k_{\rho}k_{\alpha}\epsilon_{\mu\beta}
-k_{\alpha}k_{\beta}\epsilon_{\mu\rho}
-k_{\rho}k_{\mu}\epsilon_{\alpha\beta}
+k_{\beta}k_{\mu}\epsilon_{\alpha\rho})
\end{equation}
which is identically zero, as it should be, before or after doing the harmonic coordinate transformations.

\subsection{Rotationally invariant waves}
It is worth noticing that if we start from the metric with lower indices, of the form 
\begin{equation}
g_{\mu\nu}=\eta_{\mu\nu}+\epsilon_{\mu\nu} F(k\cdot x)
\end{equation}
and impose rotational invariance in the plane perpendicular to the direction of propagation, we find an even simpler result.  Imposing of course the gauge conditions (\ref{w}) and (\ref{2})
we find, after some calculation, that the only form allowed is
\begin{equation}
g_{\mu\nu}=\eta_{\mu\nu}-ak_{\mu}k_{\nu}F
\label{metricreduced}
\end{equation}
with
\begin{equation}
g^{\mu\nu}=\eta^{\mu\nu}+ak^{\mu}k^{\nu}F
\label{inversemetricreduced}
\end{equation}
for any generic differentiable function $F=F(k\cdot x)$.  Imposing the rotational symmetry is much more restrictive than our analysis above considering the fluctuations of $g^{\mu\nu}$, as it reduces the possible metric fluctuations to only contain longitudinal waves. For such metrics, transformations of the form
\begin{equation}
\frac{\partial x'^{\mu}}{\partial x^{\nu}}=\delta^{\mu}_{\nu}-\frac{a}{2}k^{\mu}k_{\nu}F
\end{equation}
can be performed while preserving the gauge choices, and these bring the metric to the trivial form.
\section{Conclusions}
We have shown that transverse, traceless, plane fronted monochromatic gravitational waves given by the ansatz (\ref{1}) are actually conformally and coordinate equivalent to flat Minkowski space-time. This generalizes to all orders, the result found to second order in \cite{b-p} that transverse, traceless, linearized plane waves carry no energy to second order in Weyl gravity and that found in \cite{m}.  However our general result here follows from the unexpected discovery that such deviations from Minkowski space-time are actually trivial, they are just conformal and coordinate artifacts. The energy-momentum tensor of gravitational phenomena in alternative theories of gravity, and specifically here in Weyl gravity has been given in \cite{d-t}
\begin{eqnarray}
\nonumber
T_{\mu\nu}^{\mathrm{gravity}}&=
(\partial^{2}R_{\mu\nu}^{\mathrm{linear}}-\nabla^{2}R_{\mu\nu})
-\frac{1}{3}(\partial_{\mu}\partial_{\nu}R^{\mathrm{linear}}-\nabla_{\mu}\nabla_{\nu}R)
-\frac{1}{6}(\eta_{\mu\nu}\partial^{2}R^{\mathrm{linear}}-g_{\mu\nu}\nabla^{2}R)-\\
&-2R^{\alpha\beta}(R_{\beta\nu\alpha\mu}-\frac{1}{4}g_{\mu\nu}R_{\alpha\beta})
+\frac{2}{3}R(R_{\mu\nu}-\frac{1}{4}g_{\mu\nu}R).
\end{eqnarray}
For our waves, this energy-momentum tensor is identically zero.
Since our solution is valid for any strength of the fluctuation, the Weyl gravitational field equations must be valid order by order in the parameter corresponding to the strength of the fluctuation. The result in \cite{b-p} was an explicit verification of this fact for the second order terms. Here we have shown that each order vanishes exactly. How this is related to zero energy theorems in Weyl gravity is not clear \cite{b-h-s}.

We are not able to conclude from the analysis presented here that the full Weyl gravity does not contain gravitational wave phenomena.  We have shown that a single, monochromatic transverse, traceless wave does not exist to all orders in Weyl gravity, and their linearized superpositions treated to linear order are also coordinate/conformally equivalent to flat. However we have not shown that the full non-linear superpositions of waves of the form (\ref{1}) are trivial. If this were true, it would mean that Weyl gravity contains no gravitational waves of this type.  This will have important ramifications for the perturbative quantum version of the theory.  Any such analysis is done by considering fluctuations of the form Eqn. (\ref{1}) and quantizing them.  The non-existence of such waves, as proven in this paper, then requires a profound re-examination of any perturbative quantum analysis of conformal gravity.  An analysis has already been undertaken in \cite{m}, where it is shown that despite possessing no energy classically, plane-fronted waves still contribute zero-point energy quantum-mechanically. 

Other types of gravitational wave perturbations are known to exist in conformal gravity.  For example, in a recent paper \cite{m}, it is shown that perturbations of the form
\beq
g^{\mu\nu}=\eta^{\mu\nu} +B^{\mu\nu}n\cdot xF(k\cdot x)
\eeq
do carry energy and momentum.  The surprising fact that we have to report is that standard transverse, traceless monochromatic gravitational waves do not exist in conformal gravity.

\acknowledgments
We thank NSERC of Canada for financial support and the Perimeter Institute for Theoretical Physics, Waterloo, Canada for their hospitality, where this paper was finalized.  We also thank P. Mannheim, R. MacKenzie and P. R. Giri for useful discussions.  


\begin{thebibliography}{99}
%
\end{document}